\documentclass[runningheads]{llncs}
\usepackage[T1]{fontenc}
\usepackage[acronym, nopostdot]{glossaries}
\glsdisablehyper
\newacronym{FMA}{FMA}{Fused Multiply Accumulate}
\newacronym{SIMD}{SIMD}{Single Instruction Multiple Data}
\newacronym{FPGA}{FPGA}{Field Programmable Gate Array}
\newacronym{ASIC}{ASIC}{Application Specific Integrated Circuit}
\newacronym{CPU}{CPU}{Central Processing Unit}
\newacronym{GPU}{GPU}{Graphics Processing Unit}
\newacronym{HLS}{HLS}{High-Level Synthesis}
\newacronym{RTL}{RTL}{Register Transfer Level}
\newacronym{VHDL}{VHDL}{VHSIC Hardware Description Language}
\newacronym{BCPNN}{BCPNN}{Bayesian Confidence Propagation Neural Network}
\newacronym{BCP}{BCP}{Bayesian Confidence Propagation}
\newacronym{FIFO}{FIFO}{First In First Out}
\newacronym{HBM}{HBM}{High Bandwidth Memory}
\newacronym{PCIe}{PCIe}{Peripheral Component Interconnect Express}
\newacronym{AXI}{AXI}{Advanced eXtensible Interface}
\newacronym{BRAM}{BRAM}{Block Random Access Memory}
\newacronym{DMA}{DMA}{Direct Memory Access}
\newacronym{XRT}{XRT}{Xilinx Runtime library}
\newacronym{SLR}{SLR}{Super Logic Region}
\newacronym{NAISS}{NAISS}{National Academic Infrastructure for Supercomputing in Sweden}
\newacronym{DL}{DL}{Deep Learning}
\newacronym{HCU}{HCU}{hypercolumn unit}
\newacronym{DSP}{DSP}{Digital Signal Processor}
\newacronym{LUT}{LUT}{Look-Up Table}
\newacronym{TRL}{TRL}{Technology Readiness Level}
\newacronym{CU}{CU}{Compute Unit}
\newacronym{NPU}{NPU}{Neuron Processing Unit}
\newacronym{LIF}{LIF}{Leaky Integrate-and-Fire}
\newacronym{SNN}{SNN}{Spiking Neural Network}
\newacronym{AI}{AI}{Artificial intelligence}
\newacronym{URAM}{URAM}{UltraRAM}
\newacronym{RTF}{RTF}{Real-Time Factor}
\newacronym{HPC}{HPC}{High-Performance Computing}
\usepackage{comment}
\usepackage{array}
\usepackage[table]{xcolor}
\usepackage{wrapfig}
\usepackage{colortbl}
\usepackage{graphicx}
\usepackage{booktabs}   
\usepackage{pgfplots}   
\usepgfplotslibrary{groupplots}
\pgfplotsset{compat=1.18} 
\definecolor{givenbg}{RGB}{230,230,230}
\definecolor{filledbg}{RGB}{220,235,255}
\newcommand{\g}[1]{\cellcolor{givenbg}\textbf{#1}}
\newcommand{\f}[1]{\cellcolor{filledbg}#1}

\begin{document}
\title{NeuroRing: Scaling Spiking Neural Networks via Multi-FPGA Bidirectional Ring Topologies and Stream-Dataflow Architectures}
\titlerunning{NeuroRing: Scaling Spiking Neural Networks via Multi-FPGA }
%
\author{Muhammad Ihsan Al Hafiz\orcidID{0000-0002-9150-3847} \and
Artur Podobas\orcidID{0000-0001-5452-6794}}


%
\authorrunning{M. I. Al Hafiz and A. Podobas}
%
\institute{KTH Royal Institute of Technology, Stockholm, Sweden  \\
\email{\{miahafiz, podobas\}@kth.se}}
\maketitle              
\begin{abstract}
Spiking neural networks (SNNs) are a promising paradigm for energy-efficient event-driven computation, but large-scale SNN execution remains challenging because sparse spike communication and synchronization can dominate runtime. Existing solutions across CPU, GPU, ASIC, and FPGA platforms offer different trade-offs between programmability, efficiency, and scalability. To address this gap, we present NeuroRing, a modular and scalable SNN accelerator based on a stream-dataflow architecture and a bidirectional ring topology, implemented in High-Level Synthesis (HLS) on FPGAs. NeuroRing supports modular single- and multi-FPGA deployment and is compatible with existing SNN workflows through integration with the NEST simulator. We evaluate NeuroRing on the cortical microcircuit benchmark and a Sudoku constraint-satisfaction workload. Results show that NeuroRing preserves the key activity statistics of the NEST reference model, achieves faster-than-real-time execution of the full-scale cortical microcircuit with a real-time factor (RTF) of 0.83, exhibits meaningful strong and weak scaling, and provides competitive energy efficiency on two programmable FPGAs. These results position NeuroRing as a flexible and scalable platform for both neuroscience simulation and broader event-driven applications. 

\end{abstract}
\section{Introduction}

\glspl{SNN} are a promising model for energy-efficient event-driven computation and biologically realistic brain simulation, and are used in applications ranging from computational neuroscience~\cite{kurth_sub-realtime_2022} and constraint-satisfaction problems~\cite{pignari_efficient_2025} to computer vision~\cite{agarwal_spiking_2023}. SNNs are energy-efficient because of sparse spike-based communication, where neurons exchange information only when events occur. However, large-scale \gls{SNN} workloads remain challenging, particularly for networks with high connection density, such as the cortical microcircuit model~\cite{potjans_cell-type_2014}. In such workloads, spike communication and synchronization can dominate execution cost, requiring specialized architectures that exploit the characteristics of event-driven computation.

A broad range of systems has been developed to accelerate large-scale \gls{SNN} simulation, each with different trade-offs. Specialized neuromorphic hardware, such as SpiNNaker, has demonstrated real-time simulation of the full cortical microcircuit at biological time scales \cite{rhodes_real-time_2019}, but such platforms are less accessible than commodity \gls{CPU} and \gls{GPU} systems. \gls{GPU}-based approaches have also achieved near-real-time performance on similar workloads~\cite{golosio_fast_2021}, while \gls{HPC} deployments of the NEST simulator \cite{gewaltig_nest_2007} have demonstrated strong performance and straightforward scaling across compute nodes~\cite{kurth_sub-realtime_2022}. Reconfigurable hardware offers another attractive option: \gls{FPGA}-based systems have shown strong energy efficiency and flexibility \cite{szczerek_quarter_2025}, from single-node devices~\cite{lindqvist_algorithms_2024} to large \gls{FPGA} clusters~\cite{heittmann_simulating_2022,kauth_neuroaix-framework_2023}. However, existing FPGA-based SNN accelerators are often optimized for specific workloads, hardware organizations, or deployment settings. As a result, there remains a need for architectures that combine modularity, scalable multi-FPGA execution, backend-agnostic integration, and practical portability, while remaining suitable for high-performance large-scale SNN simulation.

In this work, we propose NeuroRing, a stream-dataflow \gls{SNN} architecture with a bidirectional ring topology and modular \gls{HLS} kernels for multi-\gls{FPGA} scaling. The main contributions are as follows:
\begin{enumerate}
\item We design and implement NeuroRing as a modular \gls{SNN} accelerator that supports both single- and multi-\gls{FPGA} deployment via stream-based communication and bidirectional ring routing.
\item We integrate NeuroRing with the NEST simulator through a Python-based workflow and evaluate it on two representative workloads: the cortical microcircuit for neuroscience and a Sudoku solver for constraint satisfaction.
\item We analyze the architectural trade-offs of NeuroRing through design-space exploration and evaluate its scalability under strong and weak scaling.
\item We compare NeuroRing with the Dardel supercomputer baseline, showing competitive strong-scaling performance with substantially lower power and energy consumption.
\end{enumerate}

\section{Spiking Neural Networks}

\glspl{SNN} are the third generation of artificial neural networks that encode information in the timing of spikes \cite{ghosh-dastidar_spiking_2009}. These networks are inspired by biological neurons, which use action potentials called spikes, to encode temporal information in their signals \cite{vreeken_spiking_2003}. This mechanism for transferring information can lead to energy savings in hardware implementation because it transmits and processes sparse spike events only when activity occurs, as in the human brain. 

The dynamics of spiking neurons can be described through different -- more or less biologically plausible -- models. The neuron model represents the mathematical formulation of the behavior of the neural membrane potential and ion channel conductances \cite{ghosh-dastidar_spiking_2009}. Examples of neuron models include the \gls{LIF} model, the Izhikevich model, and the Hodgkin-Huxley model. The \gls{LIF} model is a common neural model popular for its simplicity and ability to represent what is observed in vitro. Formally, the \gls{LIF} is composed of two Ordinary Differential Equations (ODEs) with a spike-generation event condition:

\begin{equation}
\label{eq:neuron_model}
\mathrm{LIF}\left\{
\begin{aligned}
&\frac{dV}{dt} =
\begin{cases}
0,\; V=V_{\mathrm{reset}}, 
& t-t_{\mathrm{last}} < \tau_{\mathrm{ref}}, \\[2pt]
\dfrac{-(V-E_{\mathrm{L}}) + (I_{\mathrm{syn}} + I_{\mathrm{DC}})R}{\tau_m},
& t-t_{\mathrm{last}} \geq \tau_{\mathrm{ref}},
\end{cases} \\[6pt]
&\frac{dI_{\mathrm{syn}}}{dt}
= - \frac{I_{\mathrm{syn}}}{\tau_{\mathrm{syn}}}
+ \sum_j w_j \delta(t - t^j), \\[4pt]
&\text{if } V(t^-) > V_{\mathrm{th}}:
\mathrm{emit\_spike}(),\;
V(t^+) = V_{\mathrm{reset}},\;
t_{\mathrm{last}} = t .
\end{aligned}
\right.
\end{equation}

The \gls{LIF} model integrates input current over time and emits a spike when the membrane potential reaches a threshold. The neuron dynamics are represented by Equation~\ref{eq:neuron_model} \cite{rhodes_real-time_2019}. The membrane potential $V$ evolves with time constant $\tau_m$ relative to the leak reversal potential (resting potential) $E_{\mathrm{L}}$, driven by the synaptic current $I_{\mathrm{syn}}$, direct current $I_{\mathrm{DC}}$, and membrane resistance $R$. The synaptic current decays with time constant $\tau_{\mathrm{syn}}$ and increases upon spike arrival, where $\delta(t-t^j)$ denotes an incoming spike at time $t^j$, and $w_j$ is the synaptic weight associated with that spike, determining the magnitude and sign of its contribution to $I_{\mathrm{syn}}$. Positive and negative weights correspond to excitatory and inhibitory contributions, respectively. When $V$ exceeds the threshold potential $V_{\mathrm{th}}$, the neuron emits a spike, resets to $V_{\mathrm{reset}}$, and updates the last spike time $t_{\mathrm{last}}$. During the refractory period $\tau_{\mathrm{ref}}$, i.e., when $t-t_{\mathrm{last}} < \tau_{\mathrm{ref}}$, the membrane potential is clamped at \(V_{\mathrm{reset}}\) and does not respond to synaptic input, while the synaptic current continues to follow its decay dynamics. Example networks using \gls{LIF} in this paper are the cortical microcircuit \cite{potjans_cell-type_2014} and the Sudoku solver \cite{j_gille_s_furber_a_rowley_script_nodate}.

\section{Related work}
A wide range of SNN accelerators has been proposed across ASIC, CPU, GPU, and FPGA platforms to improve simulation speed and energy efficiency. A common, non-trivial benchmark used to quantify accelerator performance is the cortical microcircuit model~\cite{potjans_cell-type_2014}, which has a dense (and realistic) connection density~\cite{kauth_neuroaix-framework_2023}. The cortical microcircuit benchmark is particularly challenging because its relatively large synaptic fanout makes spike communication and distribution a major design concern, unlike more structured computer-science workloads with smaller connection fanout.

In 2019, Rhodes et al.~\cite{rhodes_real-time_2019} demonstrated real-time simulation of the cortical microcircuit on SpiNNaker, a specialized neuromorphic platform. Subsequent work explored more accessible platforms. Golosio et al.~\cite{golosio_fast_2021} used a GPU-based approach and achieved near-real-time performance on a similar workload. Kurth et al.~\cite{kurth_sub-realtime_2022} later mapped the microcircuit model built in the NEST simulator to \gls{HPC} compute nodes, achieving performance beyond biological real time while also enabling straightforward scaling across supercomputer nodes.

Reconfigurable hardware has further improved the balance between flexibility and energy efficiency. Heittmann et al.~\cite{heittmann_simulating_2022} used a large FPGA cluster of 432 Xilinx Z7045 devices and achieved a \gls{RTF} of 0.25, at an energy cost of 783~nJ per synaptic event. Kauth et al.~\cite{kauth_neuroaix-framework_2023} reduced the platform scale to 35 NetFPGA SUME devices and reported an \gls{RTF} of 0.05 with 48~nJ per synaptic event. More recently, Lindqvist et al.~\cite{lindqvist_algorithms_2024} used a single-node Agilex~7 FPGA and achieved an \gls{RTF} of 0.79 with 21~nJ per synaptic event.

These studies show that FPGAs are attractive for scalable SNN simulation, but they also highlight a remaining challenge: building modular architectures that are scalable across single- and multi-device deployments and flexible with respect to both the simulator backend and the workload domain. NeuroRing addresses this gap through a stream-dataflow architecture and a bidirectional ring topology that supports modular deployment, backend-agnostic integration, and the execution of workloads from both neuroscience and constraint satisfaction.

\section{The NeuroRing Architecture}

\subsection{Stream-Dataflow Processing Core}

The NeuroRing processing core is built from the \textit{NeuroRing} and \textit{SynapseRouter} kernels, which follow a stream-dataflow design. These kernels are instantiated as compute units and paired to form one NeuroRing core. Most on-chip communication is stream-based, allowing the compute units to execute concurrently and overlap computation with communication. As shown in Fig.~\ref{fig:neuroringcore}, one NeuroRing core consists of two \glspl{CU}: the \textit{NeuroRing} \gls{CU}, and the \textit{SynapseRouter} \gls{CU}. The NeuroRing \gls{CU} contains three main processes: the \gls{NPU}, synapse-list fetch, and spike recorder. The \gls{NPU} performs the neuron-state update and spike generation. \textit{NeuroRing} supports the \gls{LIF} neuron model and Poisson generator, although other neuron models can be integrated by modifying the \gls{NPU} code. The \gls{NPU} uses an 8-lane pipelined architecture, where eight 32-bit floating-point synaptic weights are processed in parallel from a packed 256-bit input stream. Neuron-state variables, such as membrane potential and synaptic current, are stored in on-chip memory and partitioned by a factor of 8 to sustain parallel access by the processing lanes.

\begin{figure}[!t]
    \centering
    \includegraphics[width=0.9\linewidth]{fig_NeuroringCore.drawio.png}
    \caption{Architecture of the proposed NeuroRing core, showing \textbf{(left)} the overarching core architecture, and \textbf{(right)} details of the NPU and the accumulator structure.}
    \label{fig:neuroringcore}
\end{figure}

The \gls{NPU} outputs spike events, which are passed to the synapse-list fetch process. For each emitted spike, the corresponding outgoing synapse list is fetched from \gls{HBM}. Each synapse entry is encoded as a 64-bit packet containing a 32-bit synaptic weight, a 2-bit synchronization field, a 22-bit destination field, and an 8-bit delay field. These entries are transferred via 256-bit \gls{AXI} burst accesses, corresponding to four synapse packets per transaction. To reduce routing latency, the synapse list is sorted according to destination-core proximity, so that locally consumed and nearest-neighbor packets are generated first. The fetched packets are then dispatched to either the left or right ring stream based on the shorter route to the destination core, with routing determined by the destination core ID. After the synapse list of each spiking neuron has been emitted, a local synchronization token is inserted to ensure that all packets associated with the current spike are propagated before the next neuron is processed, helping prevent ring congestion and deadlock. After all spikes in the current timestep have been emitted, the \gls{NPU} generates a global synchronization token that propagates through the right ring to signal the timestep's completion across all cores. During this period, the spike recorder writes spike traces back to \gls{HBM}. The HBM interface uses one AXI4 master interface, which can be mapped to one or more \gls{HBM} channels depending on the required memory capacity. The HBM address space is divided into a spike-recording region and a synapse-list region.

The SynapseRouter is an independent \gls{CU} and runs concurrently with the NeuroRing \gls{CU}. It is activated by incoming \gls{AXI}-stream packets and internal streams. Each router receives packets from two sources: the local NeuroRing \gls{CU} and the neighboring core. Packets addressed to the current core are consumed locally and sent to the accumulator, while packets for other destinations are forwarded to the next core in the ring. Incoming packets from neighboring cores are prioritized to preserve ring progress and avoid deadlock under backpressure. Eight accumulators operate in parallel to process packets delivered by the left and right routers. Each accumulator is assigned a destination-address range based on the core capacity and core ID. Synaptic weights are accumulated in a circular buffer with 8-bit delay indexing. Although this supports up to 256 delay positions, the current implementation allocates 64 delay slots per neuron, which is sufficient for the evaluated workloads and reduces memory usage. The accumulator data are stored in \gls{URAM}; since each \gls{URAM} word can hold 72 bits, two 32-bit synaptic weights are packed into a single \gls{URAM} word to improve storage efficiency. At timestep completion, each accumulator releases the synaptic-weight value scheduled for the next timestep, packs it into the input stream, and forwards it to the \gls{NPU} to start the next timestep cycle.

\subsection{Bidirectional Ring Topology}

A bidirectional ring topology is adopted to support scalable development within and across \glspl{FPGA}. Because our design is modular, NeuroRing cores can be replicated and connected to neighboring cores via their left and right \gls{AXI}-stream ports. Within a single device, these connections form a closed bidirectional ring. Across devices, the ring is extended through high-speed serial links.

\begin{figure}[!t]
    \centering
    \includegraphics[width=0.9\linewidth]{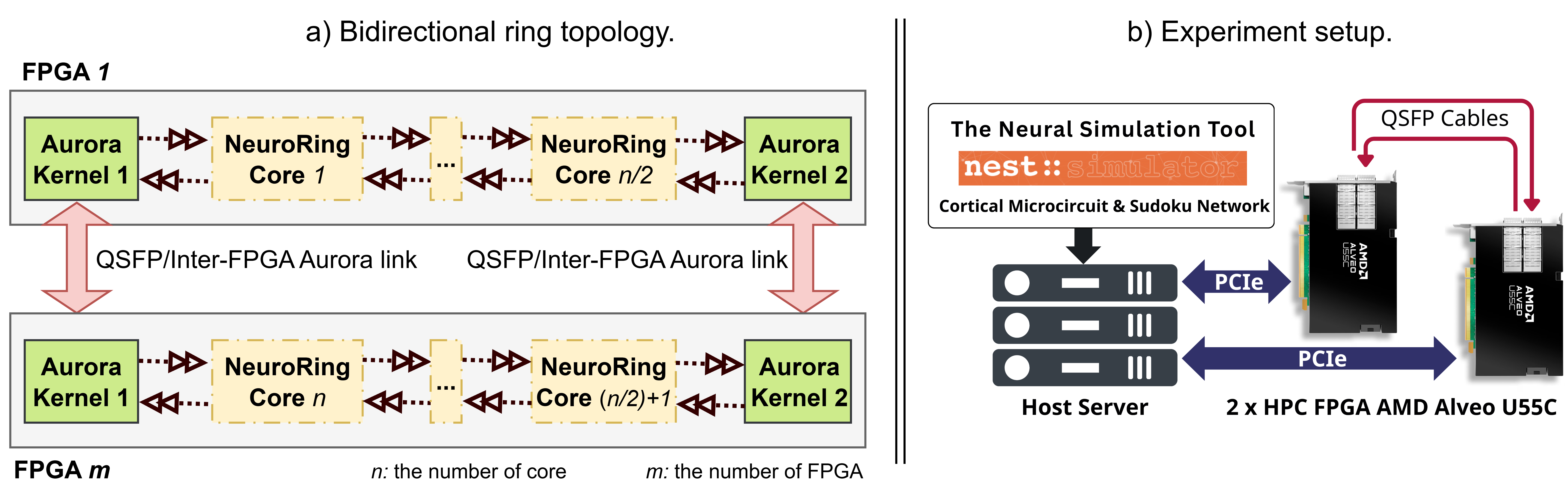}
    \caption{Bidirectional ring topology and experiment setup.}
    \label{fig:neuroring_topology}
\end{figure}

Figure~\ref{fig:neuroring_topology}a shows the system-level organization. Each device contains multiple NeuroRing cores and two Aurora kernels located at the ring boundaries. The Aurora kernels provide inter-\gls{FPGA} serial communication and extend the ring across devices via full-duplex QSFP links. This organization preserves the same left/right routing semantics used within a single \gls{FPGA} while enabling multi-device scaling. Each core is assigned a unique core ID used to determine packet routing and destination matching.

In the Vitis flow, scaling is realized by replicating NeuroRing-core instances and specifying their stream connections in the connectivity configuration file. Aurora kernels are only required for multi-\gls{FPGA} deployments. Because all target devices use the same kernel image, the generated bitstream can be reused across \glspl{FPGA}. This makes the architecture straightforward to extend from a single-device implementation to a multi-\gls{FPGA} deployment.

\subsection{Hardware Generalization}

NeuroRing is designed as a modular and backend-agnostic architecture for event-driven neural simulation. Rather than assuming a simulator-specific backend, it operates on a flattened synapse-list representation, in which each neuron stores its outgoing synaptic connections as a list of destination, delay, and synaptic weight entries. This abstraction allows NeuroRing to support different network-generation flows, provided that they can export compatible connectivity data and a supported neuron model. In this work, the network is generated using the NEST simulator \cite{gewaltig_nest_2007}, after which the synaptic connections are extracted and converted into the synapse-list representation required by NeuroRing.

The architecture can be configured for different neuron capacities per core, synapse-list sizes, and timestep settings. The number of replicated cores is limited by the available hardware resources of the target \gls{FPGA}. In addition, the neuron model can be modified by adapting the \gls{HLS} implementation of the \gls{NPU}, allowing the same architecture to support different neuron dynamics.

\section{Experimental Setup}
\label{workloads}

NeuroRing is implemented in C++ using AMD Vitis \gls{HLS} and synthesized with the Vitis 2025.1 flow for the AMD/Xilinx Alveo U55C \gls{FPGA}. Experiments were conducted on a host server equipped with two U55C cards installed via \gls{PCIe}. Inter-FPGA communication was realized through Aurora links over two QSFP cables, as shown in Fig.~\ref{fig:neuroring_topology}b. The code is publicly available at \url{https://github.com/ihsanalhafiz/NeuroRing}.

Neural workloads were generated using NEST v3.9.0, and the host runtime was implemented in Python 3.10.19 using \texttt{pyxrt} to transfer network data, launch FPGA kernels, and retrieve results. After constructing the model in the NEST simulator, we extracted the connection data using the \texttt{GetConnections} and \texttt{GetStatus} methods. The extracted connection information includes the source neuron, target neuron, synaptic weight, and synaptic delay. Synaptic weights and all neuron-state computations are represented using 32-bit floating-point precision. As a software baseline for strong scaling, we used the CPU partition of the Dardel supercomputer to run a half-scale cortical microcircuit using the same Python reference implementation. Latency was measured on the host side; it is the wall-clock simulation time unless stated otherwise. FPGA power was obtained from the average board power reported by \texttt{xrt-smi examine}, sampled during kernel execution, while CPU power on Dardel was collected using \texttt{pat\_run} and \texttt{pat\_report}. Performance is reported using the \gls{RTF}, defined as wall-clock simulation time divided by biological simulation time~\cite{lindqvist_algorithms_2024,kauth_neuroaix-framework_2023}. To evaluate performance and scalability, NeuroRing is compiled with neuron capacities of 2048, 4096, 5632, and 8192 neurons/core. Depending on the workload, the number of deployed cores ranges from 1 to 20, and both single-FPGA and two-FPGA configurations are evaluated. 

\subsection{Case Studies}
\label{case_studies}
\paragraph{\textbf{The Cortical Microcircuit}}
is a model of a unit cell of early mammalian sensory cortex that covers $1\,\mathrm{mm}^2$ of cortical surface in a spiking network. The full-scale network consists of 77{,}169 neurons that are organized into four layers of inhibitory and excitatory populations, denoted by 2/3E, 2/3I, 4E, 4I, 5E, 5I, 6E, and 6I \cite{potjans_cell-type_2014}. Neuron connections are encoded using population-specific probabilities, with a total of $\sim0.3$ billion synapses. The model can be driven by DC or Poisson-distributed external input. In this work, we use DC input for the evaluated NeuroRing and NEST reference runs. The model has become widely adopted by the computational neuroscience community for benchmarking neuroscience simulators because it involves a realistic number of connections with realistic probabilities \cite{kauth_neuroaix-framework_2023}. The synaptic fan-out averages 3{,}873 connections per neuron, with a maximum of 6{,}749. This high fan-out poses challenges for spike communication and memory bandwidth in hardware. It is simulated with a timestep of 0.1 ms, and all simulation parameters are obtained from \cite{potjans_cell-type_2014}. The cortical microcircuit is tested at Full, Half, and Quarter neuron scales with a synaptic scale factor of 1, corresponding to 77{,}169, 38{,}586, and 19{,}292 neurons, respectively. The simulation time is 10 seconds for performance evaluation.

\paragraph{\textbf{The Sudoku}} workload is derived from the NEST example network \cite{j_gille_s_furber_a_rowley_script_nodate} with adjusted parameters and implements a winner-takes-all formulation, with each Sudoku digit represented by five neurons. Digit neurons are connected to neurons representing conflicting digits in the same row, column, and 3×3 subgrid. The resulting 9$\times$9 Sudoku network contains 3645 neurons. In addition to the recurrent constraint connections, the network includes Poisson stimulation and background noise sources. Three easy Sudoku benchmark instances are used. 
\section{Results and Discussion}

\subsection{Hardware Utilization}

All kernels are synthesized with a target frequency of 300~MHz. Table~\ref{tab:hardware_util} reports the achieved frequency and hardware utilization for the evaluated NeuroRing configurations. Larger designs generally show higher utilization and lower operating frequency, reflecting a timing-closure trade-off. Smaller neuron capacities per core also raise utilization because more cores are required for the same workload. DSP usage remains relatively low because the current implementation uses the \gls{LIF} neuron model, which is less arithmetic-intensive than models such as the Izhikevich or Hodgkin--Huxley models \cite{szczerek_quarter_2025}. The Sudoku configuration has the lowest utilization, as it uses only one NeuroRing compute core and one Poisson spike-generator core.

\begin{table}[!t]
    \centering
    \caption{Hardware utilization and operating frequency of the NeuroRing architecture}
    \label{tab:hardware_util}
    \scriptsize
    \renewcommand{\arraystretch}{1}
    \setlength{\tabcolsep}{3pt}
    \begin{tabular}{@{} c c c c c c c c c c c @{}}
        \toprule
        \textbf{No} & \textbf{Core} & \textbf{Use} & \textbf{Cores} & \textbf{Setup} & \textbf{Freq.} & \textbf{LUT} & \textbf{Reg} & \textbf{BRAM} & \textbf{URAM} & \textbf{DSP} \\
        & \textbf{Capacity} & \textbf{Case} &  &  & (MHz) & (\%) & (\%) & (\%) & (\%) & (\%) \\
        \midrule
        1 & 2048 & Half CM$^{b}$ & 20 & 2 FPGAs & 268.3$^{a}$ & 49.26$^{a}$ & 37.04$^{a}$ & 44.54$^{a}$ & 26.04$^{a}$ & 18.88$^{a}$ \\
        2 & 4096 & Quarter CM$^{b}$ & 5  & 1 FPGA  & 304.1 & 29.45 & 22.07 & 20.49 & 21.88 & 9.46 \\
        3 & 4096 & Half CM$^{b}$ & 10 & 1 FPGA  & 203.5 & 49.81 & 37.94 & 31.25 & 43.75 & 18.88 \\
        4 & 4096 & Full CM$^{b}$ & 20 & 2 FPGAs & 272.4$^{a}$ & 50.24$^{a}$ & 38.46$^{a}$ & 45.29$^{a}$ & 42.71$^{a}$ & 18.88$^{a}$ \\
        5 & 5632 & Full CM$^{b}$ & 14 & 2 FPGAs & 198.7$^{a}$ & 38.59$^{a}$ & 29.26$^{a}$ & 66.67$^{a}$ & 41.25$^{a}$ & 14.01$^{a}$ \\
        6 & 8192 & Full CM$^{b}$ & 10 & 2 FPGAs & 260.0$^{a}$ & 31.13$^{a}$ & 22.87$^{a}$ & 54.71$^{a}$ & 37.50$^{a}$ & 9.46$^{a}$ \\
        7 & 4096 & Sudoku & 1$^{c}$ & 1 FPGA & 302.1 & 15.72 & 10.82 & 26.69 & 9.38 & 2.22 \\
        \bottomrule
        \multicolumn{11}{l}{\scriptsize $^{a}$Resource utilization reported per FPGA (dual-FPGA systems use identical boards).} \\
        \multicolumn{11}{l}{\scriptsize $^{b}$CM = cortical microcircuit benchmark.} \\
        \multicolumn{11}{l}{\scriptsize $^{c}$One compute core and one Poisson spike generator.}
    \end{tabular}
\end{table}

\subsection{Correctness}
\begin{figure}[!t]
    \centering
    \includegraphics[width=\linewidth]{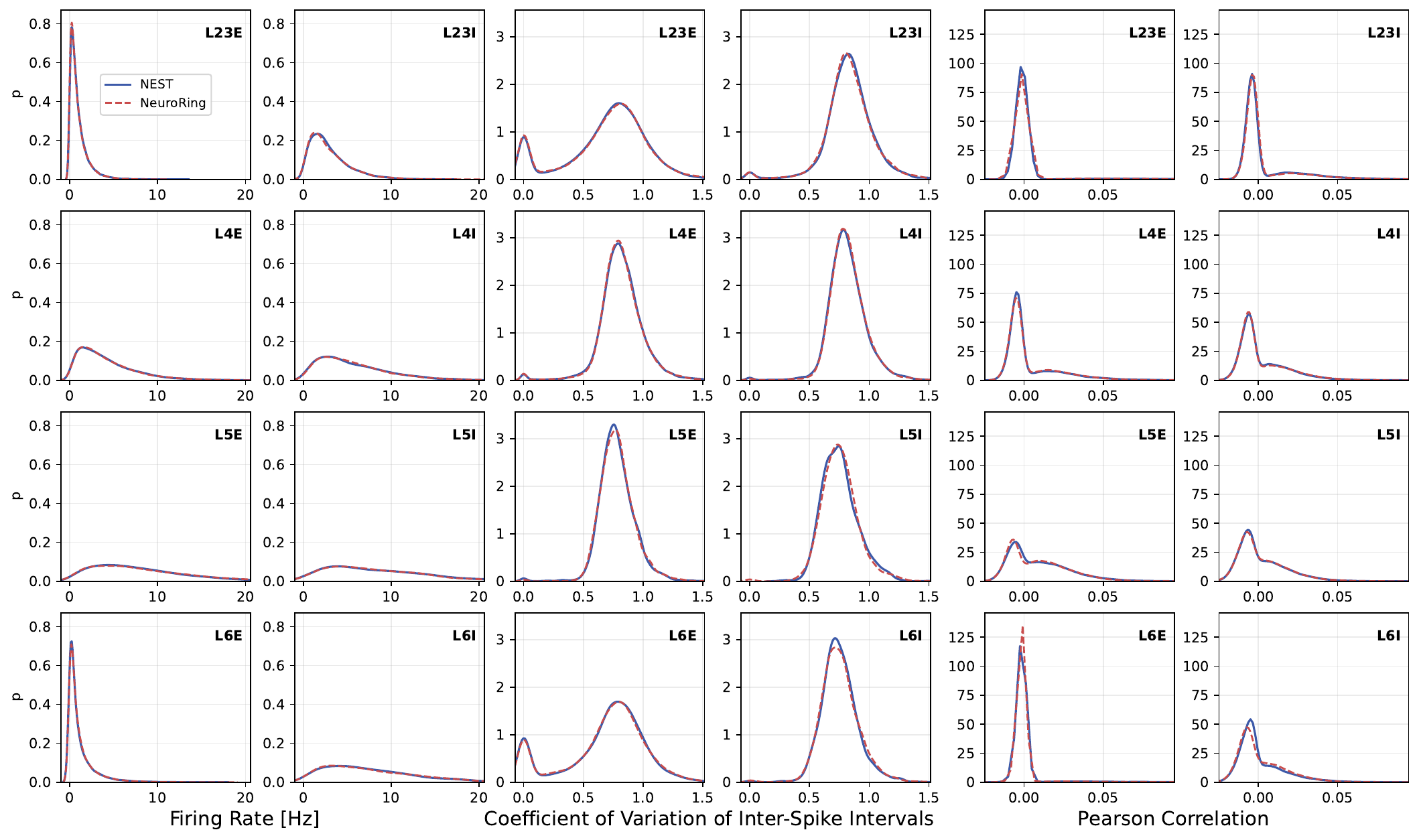}
    \caption{Layer-wise statistical comparison for the full-size cortical microcircuit.}
    \label{fig:stats}
\end{figure}

Correctness is evaluated against NEST, used here as the software reference. The comparison uses the full-scale cortical microcircuit mapped to a NeuroRing configuration with 4096 neurons/core, 20 cores, and 2 FPGAs. Validation is based on three standard spike-train statistics: firing rate, coefficient of variation (CV) of inter-spike intervals, and Pearson correlation~\cite{lindqvist_algorithms_2024,kauth_neuroaix-framework_2023,heittmann_simulating_2022}. 
Figure~\ref{fig:stats} shows that NeuroRing closely matches the NEST reference across layers L23, L4, L5, and L6 for firing rate, CV of inter-spike intervals, and Pearson correlation. The overall agreement indicates that NeuroRing preserves both the irregular spiking behavior and the correlation structure of the cortical microcircuit. Small differences remain in some tails and peak shapes, likely due to implementation-level numerical and synchronization effects. Overall, the close match in layer-wise statistics confirms that NeuroRing reproduces the essential dynamics of the NEST baseline.

\subsection{Architectural Design Space Exploration}

\begin{figure}[!t]
    \centering
    \begin{minipage}{0.50\textwidth}
        \centering
        \begin{tikzpicture}[scale=0.8]
            \begin{axis}[
                width=1.01\textwidth,
                height=4.6cm,
                title={a) RTF vs. Core Capacity},
                ylabel={RTF},
                ymin=0.5, ymax=1.5,
                xmin=0.5, xmax=3.5,
                xtick={1,2,3},
                xticklabels={8192\\(10 Cores)\\260.0 MHz, 5632\\(14 Cores)\\198.7 MHz, 4096\\(20 Cores)\\272.4 MHz},
                xticklabel style={align=center, font=\scriptsize},
                ylabel style={font=\footnotesize},
                title style={font=\footnotesize\bfseries},
                grid=major,
                grid style={dashed, gray!30}
            ]
            
            \addplot[
                color=blue,
                mark=square*,
                thick,
                nodes near coords,
                point meta=y,
                every node near coord/.style={font=\scriptsize, yshift=3pt}
            ] coordinates {
                (1, 1.11)
                (2, 1.04)
                (3, 0.83)
            };
            \end{axis}
        \end{tikzpicture}
    \end{minipage}\hfill
    \begin{minipage}{0.50\textwidth}
        \centering
        \begin{tikzpicture}[scale=0.8]
            \begin{axis}[
                width=1.01\textwidth,
                height=4.6cm,
                title={b) Power \& Energy Efficiency},
                ylabel={Total Power (W)},
                ymin=0, ymax=150,
                xmin=0.5, xmax=3.5,
                xtick={1,2,3},
                xticklabels={8192\\(10 Cores)\\260.0 MHz, 5632\\(14 Cores)\\198.7 MHz, 4096\\(20 Cores)\\272.4 MHz},
                xticklabel style={align=center, font=\scriptsize},
                ylabel style={font=\footnotesize, text=violet!80!black},
                title style={font=\footnotesize\bfseries},
                axis y line*=left,
                grid=major,
                ybar,
                bar width=10pt
            ]
            \addplot[
                fill=violet!40,
                draw=violet!80,
                nodes near coords,
                point meta=y,
                every node near coord/.style={
                    font=\scriptsize,
                    yshift=1pt,
                    text=violet!80!black
                }
            ] coordinates {
                (1, 70.24)
                (2, 69.89)
                (3, 83.08)
            };
            \end{axis}
            
            \begin{axis}[
                width=\textwidth,
                height=4.6cm,
                ylabel={Energy / Synaptic Event (nJ)},
                ymin=0, ymax=100,
                xmin=0.5, xmax=3.5,
                xtick=\empty,
                ylabel style={font=\footnotesize, text=orange!90!black},
                axis y line*=right,
                axis x line=none
            ]
            \addplot[
                color=orange!90!black,
                mark=*,
                thick,
                nodes near coords,
                point meta=y,
                every node near coord/.style={font=\scriptsize, yshift=3pt}
            ] coordinates {
                (1, 82)
                (2, 75)
                (3, 73)
            };
            \end{axis}
        \end{tikzpicture}
    \end{minipage}
    \caption{Impact of core capacity variation on RTF, Power, and Energy}
    \label{fig:design_space_exploration}
\end{figure}
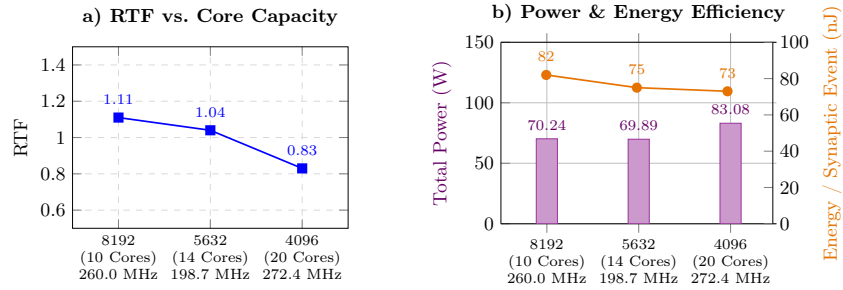

This experiment evaluates how neuron capacity per core affects RTF, power, and energy efficiency for the full-size cortical microcircuit on two FPGAs. As shown in Figure \ref{fig:design_space_exploration}a, reducing the capacity from 8192 to 4096 neurons/core improves the RTF from 1.11 to 0.83 because the workload is distributed over more cores. The 5632-neuron/core case deviates from a linear trend due to its lower post-implementation frequency of 198.7 MHz. Figure~\ref{fig:design_space_exploration}b shows that total power remains around 70 W for the 8192- and 5632-neuron/core cases, and increases to 83.08 W for the 4096-neuron/core case, where more cores are instantiated. However, the energy per synaptic event decreases from 82 nJ to 73 nJ.
Overall, these results show that reducing the neuron capacity per core improves both real-time performance and energy per synaptic event, at the cost of increased total power. These results suggest that, for this workload and platform, smaller core capacities are favorable when the primary objective is real-time performance and energy per synaptic event, provided that sufficient hardware resources are available.
\subsection{System Scalability}
\paragraph{\textbf{Strong Scaling}}
evaluates how performance changes when a fixed workload is distributed over more hardware resources. Here, the half-size cortical microcircuit is used as the fixed workload and executed on both the NeuroRing and the Dardel CPU platform under single-unit and dual-unit configurations. For NeuroRing, the single-unit configuration uses 10 cores with 4096 neurons/core on 1 FPGA, while the dual-unit configuration uses 20 cores with 2048 neurons/core on 2 FPGAs. For Dardel, the single-unit and dual-unit configurations use 1 and 2 CPU nodes, respectively, with 2 ntasks per node and 32 CPUs per task.

\begin{figure*}[!t]
    \centering
    \begin{tikzpicture}[scale=0.82]
    \begin{groupplot}[
        group style={
            group size=3 by 1,
            horizontal sep=1.5cm
        },
        width=0.4\textwidth,
        height=4.6cm,
        ymin=0,
        xtick={1,2},
        xticklabels={Single\\Unit, Dual\\Unit},
        xticklabel style={align=center, font=\scriptsize},
        grid=major,
        grid style={dashed, gray!30},
        enlarge x limits=0.5,
        legend style={
            at={(0.5,1.22)},
            anchor=south,
            legend columns=4,
            font=\scriptsize,
            draw=none
        },
        legend cell align=left,
        title style={font=\footnotesize\bfseries},
        ylabel style={font=\footnotesize}
    ]

    \nextgroupplot[
        title={a) RTF},
        ylabel={RTF},
        ymin=0, ymax=1
    ]
    \addplot[
        color=blue!80!black,
        mark=square*,
        thick,
        nodes near coords,
        point meta=y,
        every node near coord/.style={
            font=\scriptsize,
            anchor=south,
            xshift=8pt,
            yshift=2pt
        }
    ] coordinates {(1,0.73) (2,0.51)};
    
    \addplot[
        color=orange!90!black,
        mark=*,
        thick,
        nodes near coords,
        point meta=y,
        every node near coord/.style={
            font=\scriptsize,
            anchor=south,
            xshift=8pt,
            yshift=1pt
        }
    ] coordinates {(1,0.43) (2,0.36)};
    
    \legend{NeuroRing, Dardel CPU}

    \nextgroupplot[
        title={b) Power},
        ylabel={Total Power (W)},
        ymin=0, ymax=800
    ]
    \addplot[
        color=blue!80!black,
        mark=square*,
        thick,
        nodes near coords,
        point meta=y,
        every node near coord/.style={
            font=\scriptsize,
            anchor=south,
            xshift=-8pt,
            yshift=2pt
        }
    ] coordinates {(1,32.71) (2,79.17)};
    
    \addplot[
        color=orange!90!black,
        mark=*,
        thick,
        nodes near coords,
        point meta=y,
        every node near coord/.style={
            font=\scriptsize,
            anchor=south,
            xshift=-8pt,
            yshift=2pt
        }
    ] coordinates {(1,314.10) (2,587.60)};

    \nextgroupplot[
        title={c) Energy per Synaptic Event},
        ylabel={Energy/Syn.Event (nJ)},
        ymin=0, ymax=600
    ]
    \addplot[
        color=blue!80!black,
        mark=square*,
        thick,
        nodes near coords,
        point meta=y,
        every node near coord/.style={
            font=\scriptsize,
            anchor=south,
            xshift=-6pt,
            yshift=2pt
        }
    ] coordinates {(1,49) (2,84)};
    
    \addplot[
        color=orange!90!black,
        mark=*,
        thick,
        nodes near coords,
        point meta=y,
        every node near coord/.style={
            font=\scriptsize,
            anchor=south,
            xshift=-6pt,
            yshift=2pt
        }
    ] coordinates {(1,275) (2,428)};

    \end{groupplot}
    \end{tikzpicture}
    \caption{Strong-scaling comparison for the Half cortical microcircuit between NeuroRing and the Dardel CPU system.}
    \label{fig:neuroring_vs_dardel}
\end{figure*}
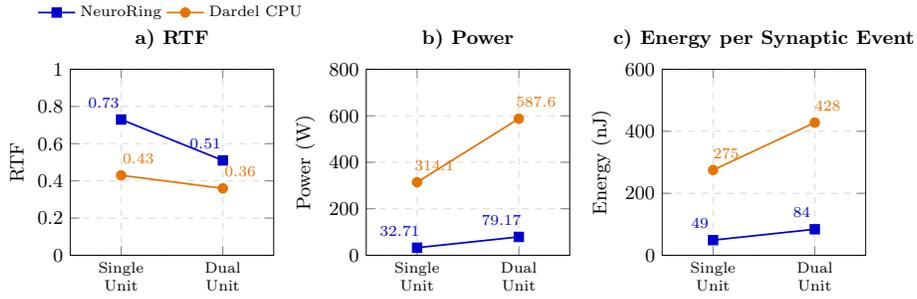

Figure~\ref{fig:neuroring_vs_dardel}a shows that scaling NeuroRing from a single-unit to a dual-unit configuration reduces the RTF from 0.73 to 0.51, or a speedup of about 1.43$\times$. For Dardel, the RTF decreases from 0.43 to 0.36, giving a smaller speedup of about 1.19$\times$. In both cases, the speedup is sublinear, indicating that communication and synchronization overhead limit ideal strong scaling. For NeuroRing, this overhead mainly comes from the longer ring and inter-FPGA communication, while for Dardel, it is associated with inter-node MPI communication.

Figures~\ref{fig:neuroring_vs_dardel}b and~\ref{fig:neuroring_vs_dardel}c show that the reduction in runtime is accompanied by higher power and energy cost on both platforms. For NeuroRing, total power increases from 32.71~W to 79.17~W, and energy per synaptic event rises from 49~nJ to 84~nJ. Dardel shows the same trend, increasing from 314.1~W to 587.6~W and from 275~nJ to 428~nJ. Although Dardel achieves lower absolute RTF in this half-size experiment, NeuroRing provides a comparable scaling trend at substantially lower power and energy, highlighting its potential as an energy-efficient platform.

\paragraph{\textbf{Weak Scaling}} evaluates performance when workload size and resources increase proportionally. Using 4096 neurons/core, the Quarter, Half, and Full cortical microcircuit workloads are mapped to 5, 10, and 20 cores.

\begin{figure}[!t]
    \centering
    \begin{minipage}{0.5\textwidth}
        \centering
        \begin{tikzpicture}[scale=0.8]
            \begin{axis}[
                width=\textwidth,
                height=4.6cm,
                title={a) Weak Scaling RTF},
                ylabel={RTF},
                ymin=0.3, ymax=1, 
                xmin=0.5, xmax=3.5,
                xtick={1,2,3},
                xticklabels={Quarter\\(5 Cores)\\(1 FPGA)\\304.1 MHz, Half\\(10 Cores)\\(1 FPGA)\\203.5 MHz, Full\\(20 Cores)\\(2 FPGAs)\\272.4 MHz},
                xticklabel style={align=center, font=\scriptsize}, 
                ylabel style={font=\footnotesize},
                title style={font=\footnotesize\bfseries},
                grid=major,
                grid style={dashed, gray!30}
            ]
            
            \addplot[color=blue, mark=square*, thick,
                nodes near coords,
                point meta=y,
                every node near coord/.style={font=\scriptsize, yshift=3pt}
            ] coordinates {
                (1, 0.52) 
                (2, 0.73) 
                (3, 0.83) 
            };
            
            \draw[dashed, thick, red] (axis cs: 2.5, 0) -- (axis cs: 2.5, 10);
            \node[anchor=south west, font=\scriptsize, text=red, align=center] 
                at (axis cs:2.5, 0.4) {Hardware\\Shift to\\2 FPGAs};
            
            \end{axis}
        \end{tikzpicture}
    \end{minipage}\hfill
    \begin{minipage}{0.5\textwidth}
        \centering
        \begin{tikzpicture}[scale=0.8]
            \begin{axis}[
                width=\textwidth,
                height=4.6cm,
                title={b) Power \& Energy Efficiency},
                ylabel={Total Power (W)},
                ymin=0, ymax=150, 
                xmin=0.5, xmax=3.5,
                xtick={1,2,3},
                xticklabels={Quarter\\(5 Cores)\\(1 FPGA)\\304.1 MHz, Half\\(10 Cores)\\(1 FPGA)\\203.5 MHz, Full\\(20 Cores)\\(2 FPGAs)\\272.4 MHz},
                xticklabel style={align=center, font=\scriptsize},
                ylabel style={font=\footnotesize, font=\color{violet}},
                title style={font=\footnotesize\bfseries},
                axis y line*=left, 
                grid=major,
                ybar, bar width=10pt
            ]
            \addplot[fill=violet!40, draw=violet!80,
                nodes near coords,
                point meta=y,
                every node near coord/.style={font=\scriptsize, yshift=1pt, text=violet!80!black }
            ] coordinates {
                (1, 27.32) 
                (2, 32.71) 
                (3, 83.08) 
            };
            
            \draw[dashed, thick, red] (axis cs: 2.5, 0) -- (axis cs: 2.5, 150);
            \end{axis}
            
            \begin{axis}[
                width=\textwidth,
                height=4.6cm,
                ylabel={Energy / Synaptic Event ($n$J)},
                ymin=0, ymax=100, 
                xmin=0.5, xmax=3.5,
                xtick=\empty, 
                ylabel style={font=\footnotesize, font=\color{orange!90!black}},
                axis y line*=right, 
                axis x line=none 
            ]
            \addplot[color=orange, mark=*, thick,
                nodes near coords,
                point meta=y,
                every node near coord/.style={font=\scriptsize, yshift=3pt}
            ] coordinates {
                (1, 56) 
                (2, 49) 
                (3, 73) 
            };
            \end{axis}
        \end{tikzpicture}
    \end{minipage}
    \caption{Weak-scaling performance of the NeuroRing architecture (4096 neurons/core).}
    \label{fig:scaling_performance}
\end{figure}
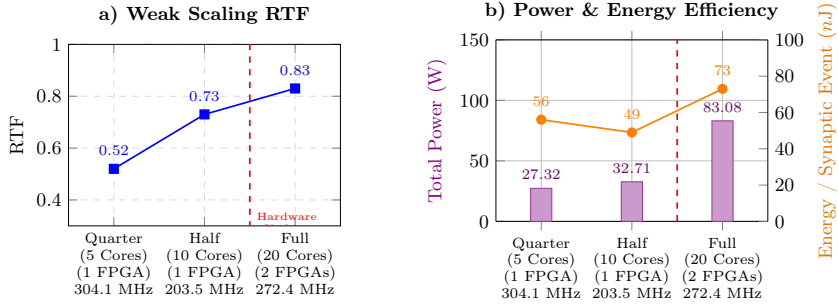

Ideally, the RTF would remain constant. Instead, Figure \ref{fig:scaling_performance}a shows that the RTF increases from 0.52 for the Quarter workload to 0.73 for the Half workload and 0.83 for the Full workload, indicating a deviation from ideal weak scaling. The degradation from Quarter to Half occurs even within a single FPGA and is likely influenced by the lower post-implementation frequency of the Half configuration (203.5~MHz) compared with the Quarter configuration (304.1~MHz). The Full configuration reaches a higher frequency (272.4~MHz), but its execution is distributed across two FPGAs, which introduces additional inter-FPGA communication and synchronization overhead. As a result, the larger configuration still shows a higher RTF. 
Figure~\ref{fig:scaling_performance}b shows that total power increases with scale, rising from 27.32~W for Quarter to 32.71~W for Half and 83.08~W for Full. The increase is particularly pronounced when moving to the two-FPGA configuration. In terms of energy per synaptic event, the Half configuration gives the best result at 49~nJ, compared with 56~nJ for Quarter and 73~nJ for Full. 
Overall, proportional resource growth is applied, but weak-scaling efficiency degrades due to lower frequency and inter-FPGA overhead.

\subsection{State-of-the-art Comparison}

To position NeuroRing with respect to prior SNN simulation platforms, Table~\ref{tab:sota_comparison} compares the full-size cortical microcircuit use case across FPGA, GPU, CPU, and ASIC-based systems. For NeuroRing, we report the configuration with 20 cores, 4096 neurons/core, deployed on 2 AMD/Xilinx Alveo U55C FPGAs.

\begin{table}[!t]
\centering
\caption{State-of-the-art Comparison for the full-size cortical microcircuit use case.}
\label{tab:sota_comparison}
\scriptsize
\renewcommand{\arraystretch}{1}
\setlength{\tabcolsep}{5pt}
\begin{tabular}{@{} l l c c c @{}}
\toprule
\textbf{System} & \textbf{Hardware} & \textbf{Platform} & \textbf{RTF} & \textbf{Energy/Syn.Ev.(nJ)} \\
\midrule
Fast SNN FPGA \cite{lindqvist_algorithms_2024} & 1 Agilex 7 & FPGA & 0.79 & 21 \\
neuroAIx \cite{kauth_neuroaix-framework_2023} & 35 NetFPGA SUME & FPGA & 0.05 & 48 \\
IBM INC-3000 \cite{heittmann_simulating_2022} & 432 Xilinx XCZ7045 & FPGA & 0.25 & 783 \\
NeuronGPU \cite{golosio_fast_2021} & 1 GeForce RTX 2080 Ti & GPU & 1.06 & 180 \\
NEST \cite{kurth_sub-realtime_2022} & 2 AMD EPYC Rome & CPU & 0.53 & 480 \\
SpiNNaker \cite{rhodes_real-time_2019} & 318 SpiNNaker chips & ASIC & 1.00 & 600 \\
\midrule
\textbf{NeuroRing (4096)} & 2 AMD/Xilinx Alveo U55C & FPGA & 0.83 & 73 \\
\bottomrule
\end{tabular}
\end{table}

Table~\ref{tab:sota_comparison} compares NeuroRing with prior full-size cortical microcircuit implementations across FPGA, GPU, CPU, and ASIC platforms. NeuroRing achieves faster-than-real-time execution with an RTF of 0.83 using two AMD/Xilinx Alveo U55C FPGAs. Although it does not provide the lowest reported energy per synaptic event among FPGA systems, it is more energy-efficient than the listed GPU and CPU baselines and improves upon the large SpiNNaker \cite{rhodes_real-time_2019} and IBM INC-3000 systems \cite{heittmann_simulating_2022} in energy per synaptic event.
NeuroRing’s main distinction is its modular HLS-based implementation on commercial data-center FPGAs, together with a Python/NEST workflow that enables existing NEST models to be mapped to the FPGA backend after model construction.
Overall, NeuroRing is positioned as a programmable and scalable FPGA-based SNN simulation platform that balances real-time performance, energy efficiency, and architectural flexibility.

\subsection{Beyond Neuroscience: Constraint Satisfaction Problem}

To demonstrate applicability beyond neuroscience, NeuroRing is evaluated on three Sudoku puzzles formulated as SNN winner-takes-all networks, following the NEST example in Section \ref{case_studies}. All instances use one NeuroRing compute core with 4096 neurons/core and one auxiliary Poisson spike generator core. The same parameter set is used for all three instances, including 200~Hz stimulus and noise rates, inhibitory weight $-100$, stimulus/noise weights 200, synaptic delay 1.0~ms, and \gls{LIF} neuron parameters $dt=0.1$~ms, $C_m=250$~pF, $I_e=200$~pA, $\tau_m=20$~ms, $t_{ref}=2$~ms, $\tau_{syn,ex}=\tau_{syn,in}=5$~ms, $V_{reset}=-70$~mV, $E_L=-65$~mV, $V_{th}=-50$~mV, and $V_m \sim \mathcal{U}(-65,-55)$~mV. Different puzzles may require retuning input-rate and synaptic-weight parameters; we leave systematic transferability analysis to future work. The solution is decoded from the digit population with the highest spike count in each cell.

\begin{figure*}[!t]
\centering
\setlength{\tabcolsep}{1.5pt}
\renewcommand{\arraystretch}{1}

\begin{minipage}{0.3\textwidth}
\centering
\scriptsize
\textbf{Sudoku 1}\\[1mm]
\begin{tabular}{|c|c|c||c|c|c||c|c|c|}
\hline
\f{6} & \g{5} & \g{8} & \f{9} & \g{3} & \f{1} & \f{4} & \g{2} & \f{7} \\ \hline
\g{4} & \f{3} & \g{2} & \f{6} & \f{7} & \f{8} & \g{9} & \f{1} & \g{5} \\ \hline
\f{9} & \f{1} & \g{7} & \f{2} & \f{4} & \f{5} & \g{6} & \g{8} & \f{3} \\ \hline\hline
\g{2} & \g{9} & \f{6} & \f{3} & \g{5} & \g{4} & \f{8} & \g{7} & \f{1} \\ \hline
\g{5} & \f{8} & \f{1} & \f{7} & \g{6} & \g{2} & \f{3} & \f{4} & \f{9} \\ \hline
\f{7} & \f{4} & \g{3} & \g{8} & \g{1} & \f{9} & \g{2} & \g{5} & \f{6} \\ \hline\hline
\g{1} & \f{2} & \g{9} & \f{5} & \f{8} & \g{3} & \f{7} & \g{6} & \g{4} \\ \hline
\g{8} & \g{6} & \g{5} & \g{4} & \g{9} & \f{7} & \g{1} & \g{3} & \f{2} \\ \hline
\f{3} & \g{7} & \f{4} & \f{1} & \f{2} & \g{6} & \f{5} & \f{9} & \f{8} \\ \hline
\end{tabular}
\end{minipage}\hfill
\begin{minipage}{0.3\textwidth}
\centering
\scriptsize
\textbf{Sudoku 2}\\[1mm]
\begin{tabular}{|c|c|c||c|c|c||c|c|c|}
\hline
\f{2} & \g{5} & \f{4} & \f{6} & \f{8} & \f{3} & \g{1} & \g{9} & \f{7} \\ \hline
\f{6} & \f{8} & \f{7} & \f{9} & \f{1} & \f{5} & \f{3} & \g{4} & \g{2} \\ \hline
\g{9} & \g{1} & \f{3} & \f{4} & \g{2} & \g{7} & \g{5} & \g{6} & \g{8} \\ \hline\hline
\g{3} & \g{4} & \g{5} & \f{8} & \f{7} & \g{1} & \g{9} & \f{2} & \g{6} \\ \hline
\g{7} & \f{2} & \f{6} & \g{3} & \g{4} & \f{9} & \f{8} & \f{5} & \f{1} \\ \hline
\g{8} & \g{9} & \f{1} & \g{2} & \f{5} & \g{6} & \f{4} & \f{7} & \g{3} \\ \hline\hline
\f{5} & \f{3} & \g{8} & \g{7} & \f{6} & \f{4} & \g{2} & \g{1} & \f{9} \\ \hline
\g{1} & \g{6} & \f{2} & \f{5} & \f{9} & \g{8} & \f{7} & \f{3} & \g{4} \\ \hline
\f{4} & \f{7} & \f{9} & \g{1} & \f{3} & \f{2} & \g{6} & \g{8} & \g{5} \\ \hline
\end{tabular}
\end{minipage}\hfill
\begin{minipage}{0.3\textwidth}
\centering
\scriptsize
\textbf{Sudoku 3}\\[1mm]
\begin{tabular}{|c|c|c||c|c|c||c|c|c|}
\hline
\f{4} & \f{9} & \g{6} & \f{7} & \f{1} & \f{5} & \g{8} & \f{3} & \g{2} \\ \hline
\g{7} & \f{5} & \f{3} & \g{4} & \g{2} & \g{8} & \f{1} & \g{9} & \g{6} \\ \hline
\g{2} & \g{1} & \f{8} & \f{6} & \g{3} & \f{9} & \g{7} & \f{4} & \g{5} \\ \hline\hline
\f{5} & \g{3} & \g{1} & \f{2} & \f{6} & \f{7} & \g{9} & \g{8} & \f{4} \\ \hline
\f{6} & \f{4} & \f{9} & \g{1} & \f{8} & \f{3} & \f{2} & \f{5} & \g{7} \\ \hline
\g{8} & \g{2} & \f{7} & \g{9} & \g{5} & \f{4} & \f{6} & \f{1} & \g{3} \\ \hline\hline
\g{3} & \f{7} & \f{4} & \f{8} & \f{9} & \g{2} & \f{5} & \g{6} & \f{1} \\ \hline
\f{1} & \g{8} & \g{5} & \f{3} & \g{7} & \g{6} & \f{4} & \f{2} & \f{9} \\ \hline
\g{9} & \f{6} & \g{2} & \g{5} & \f{4} & \g{1} & \g{3} & \f{7} & \g{8} \\ \hline
\end{tabular}
\end{minipage}
\par\medskip
\begin{minipage}{0.95\textwidth}
\centering
\scriptsize
\textbf{Performance summary:} Sim.Time = 0.5~s, End-to-End Latency = 0.53--0.57~s,\\
SNN Exec. Latency = 0.06--0.07~s, Power $\approx$ 20.8~W, Energy = 109--110~nJ/synaptic event.
\end{minipage}
\caption{Three Sudoku benchmarks. Gray cells: given puzzle. Blue cells: solver solution.}
\label{fig:sudoku_benchmarks}
\end{figure*}

Figure~\ref{fig:sudoku_benchmarks} shows that all three puzzles are solved correctly. The results show stable performance across the instances, with end-to-end latency of 0.53--0.57~s, SNN execution latency of 0.06--0.07~s, and power of about 20.8~W. The higher energy per synaptic event compared with the cortical microcircuit case is due to the smaller number of synaptic events, so the fixed platform power is distributed over fewer spike-processing events. These results show that NeuroRing can support constraint satisfaction problems beyond neuroscience benchmarks on a compact single-FPGA configuration.

\section{Conclusion}

This paper presented NeuroRing, a modular and scalable multi-FPGA architecture for spiking neural network simulation based on a bidirectional ring topology and stream-dataflow execution in \gls{HLS}. NeuroRing supports configurable neuron capacity per core, scales across single- and multi-FPGA deployments, and preserves the key statistics of the NEST reference model. Results showed faster-than-real-time execution of the full-size cortical microcircuit, meaningful strong and weak scaling, and competitive energy efficiency using only two programmable FPGAs. The Sudoku case study further demonstrated NeuroRing’s applicability to constraint-satisfaction workloads beyond neuroscience. These results highlight the potential of NeuroRing as a flexible and efficient architecture for scalable \gls{SNN} simulation and related event-driven workloads. Future work will focus on extending NeuroRing with reconfigurable neuron models and on-chip learning capabilities, enabling support for more diverse \gls{SNN} workloads and adaptive neural dynamics.

\begin{credits}
\subsubsection{\ackname} This work was funded by the European Commission Directorate-General for Communications Networks, Content and Technology, grant no. 101135809 (EXTRA-BRAIN), the Swedish Research Council grant no. 2021-04579 (Building Digital Brains), and the Swedish e-Science Research Centre (SeRC). The computations were enabled by resources provided by the National Academic Infrastructure for Supercomputing in Sweden (NAISS), partially funded by the Swedish Research Council through grant agreement no. 2022-06725.

\subsubsection{\discintname} The authors have no competing interests to declare that are relevant to the content of this article.
\end{credits}

%
%
%
\bibliographystyle{splncs04}
\bibliography{references_refine}
\end{document}